\documentclass[aps,superscriptaddress,prb,amsmath,amssymb,reprint]{revtex4-1}
\usepackage{color}
\usepackage{epsfig}
\usepackage{hyperref}
\usepackage{graphicx}
\usepackage{dcolumn}
\usepackage{bm}

\begin{document}

\title{Nuclear spin dynamics influenced and detected by electron spin polarization in CdTe/(Cd,Mg)Te quantum wells}

\author{E.~Evers}
\email[Corresponding author: ]{eiko.evers@tu-dortmund.de}
\affiliation{Experimentelle Physik 2, Technische Universit\"at Dortmund, D-44221 Dortmund, Germany}
\author{T.~Kazimierczuk}
\affiliation{Experimentelle Physik 2, Technische Universit\"at Dortmund, D-44221 Dortmund, Germany}
\affiliation{Institute of Experimental Physics, Faculty of Physics, University of Warsaw, PL-02093 Warsaw, Poland}
\author{F.~Mertens}
\altaffiliation[Present address: ]{Experimentelle Physik 6, Technische Universit\"at Dortmund, D-44221 Dortmund, Germany.}
\affiliation{Experimentelle Physik 2, Technische Universit\"at Dortmund, D-44221 Dortmund, Germany}
\author{D.~R. Yakovlev}
\affiliation{Experimentelle Physik 2, Technische Universit\"at Dortmund, D-44221 Dortmund, Germany}
\affiliation{Ioffe Institute, Russian Academy of Sciences, 194021 St.\,Petersburg, Russia}
\author{G.~Karczewski}
\affiliation{Institute of Physics, Polish Academy of Sciences, PL-02668 Warsaw, Poland}
\author{T.~Wojtowicz}
\affiliation{Institute of Physics, Polish Academy of Sciences, PL-02668 Warsaw, Poland}
\affiliation{International Research Centre MagTop, PL-02668 Warsaw, Poland}
\author{J.~Kossut}
\affiliation{Institute of Physics, Polish Academy of Sciences, PL-02668 Warsaw, Poland}
\author{M.~Bayer}
\affiliation{Experimentelle Physik 2, Technische Universit\"at Dortmund, 44221 Dortmund, Germany}
\affiliation{Ioffe Institute, Russian Academy of Sciences, 194021 St.\,Petersburg, Russia}
\author{A.~Greilich}
\affiliation{Experimentelle Physik 2, Technische Universit\"at Dortmund, 44221 Dortmund, Germany}

\begin{abstract}
    Nuclear spin coherence and relaxation dynamics of all constituent isotopes of an n-doped CdTe/(Cd,Mg)Te quantum well structure are studied employing optically detected nuclear magnetic resonance. Using time-resolved pump-probe Faraday ellipticity, we generate and detect the coherent spin dynamics of the resident electrons. The photogenerated electron spin polarization is transferred into the nuclear spin system, which becomes polarized and acts back on the electron spins as the Overhauser field. Under the influence of resonant radio frequency pulses, we trace the coherent spin dynamics of the nuclear isotopes $^{111}$Cd, $^{113}$Cd, and $^{125}$Te. We measure nuclear Rabi oscillations, the inhomogeneous dephasing time $T_2^*$, the spin coherence time $T_2$, and the longitudinal relaxation time $T_1$. Furthermore, we investigate the influence of the laser excitation and the corresponding electron spin polarization on the nuclear spin relaxation time and find a weak extension of this time induced by interaction with the electron spins.
\end{abstract}

  \maketitle

  \section{Introduction}
  The nuclear spin system of semiconductors is of particular interest as it exhibits a long coherence time as a result of a reduced interaction with the surrounding. Consequently, the direct manipulation of nuclear spins is only possible in a slow manner and, as an additional drawback, exhibits only a very low light-matter interaction. Electron spins, however, can be manipulated much more efficiently by light~\cite{RamsayReview}. Their spin acts on the nuclear spin system allowing for an electron-mediated optical nuclear spin polarization, both in bulk semiconductors~\cite{Lampel1968,Ekimov1972,OptOrientation} and in low-dimensional structures~\cite{Flinn1990,Barrett1994,UrbaszekNuclReview13,Dyakonov2017}. The nuclear spin system, in turn, is a prominent source for the electron spin dephasing~\cite{Luke_PRL102} so that mapping out the nuclear spin dynamics is a key for understanding the electron spin time evolution. Letting the electron spin detect the nuclear spin dynamics offers the possibility to study a small number of nuclear spins in the localization volume of electrons in a quantum well (QW) structure~\cite{Flinn1990}. Additionally, one might get insight into the dynamics of different isotopes. At low magnetic fields, optically detected nuclear magnetic resonance (ODNMR) can be applied by measuring the photoluminescence  polarization~\cite{Paget1977}, whereas, at higher fields, time-resolved Faraday rotation using a pump-probe technique can be carried out~\cite{PoggioAwschalom2005,SanadaPRL2006}. Furthermore, the use of radio-frequency (RF) pulses enables one to study the nuclear spin dephasing and coherence times~\cite{SanadaPRL2006,ChekhovichNatCommun2015}. In this letter, we report on a series of pulsed ODNMR experiments in a low perturbation regime. In contrast to Ref.~[\onlinecite{SanadaPRL2006}], where GaAs/(Al,Ga)As QWs were investigated, we observe the dynamics of a small nuclear polarization ($\leq 6 \, \%$) in the absence of a pumped electron spin polarization. Furthermore, in CdTe/(Cd,Mg)Te QW, all isotopes with non zero nuclear spins exhibit a total spin of 1/2, while in GaAs/(Al,Ga)As all isotopes have spins 3/2. We can, therefore, exclude the nuclear quadrupole-induced spin relaxation, which plays an important role in the nuclear spin dynamics~\cite{UrbaszekNuclReview13}. Additionally, we investigate the impact of the electron spin polarization on the nuclear relaxation dynamics.

\section{Experimental details}
\begin{figure}[t]
  \includegraphics[width=\columnwidth]{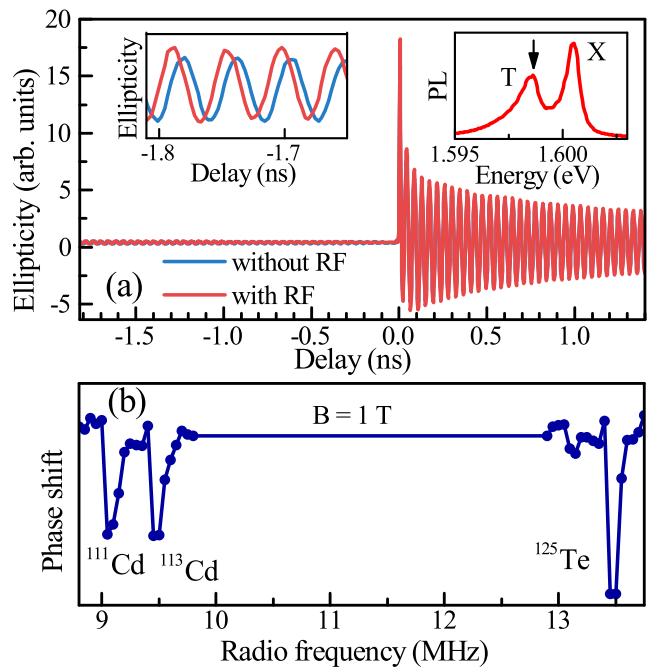}
  \caption{(a) Dynamics of the Faraday ellipticity of the probe beam polarization after passing through the sample versus time delay relative to the pump pulse measured at $B=1 \,$T. The left inset presents a zoom of the data around a delay of $-1.7\,$ns (i.e. $11.5\,$ns after the previous pump pulse). The red curve corresponds to the measurement in the presence of RF, affecting the nuclear polarization. The right inset shows the photoluminescence spectrum (PL) exhibiting trion (T) and exciton (X) emission lines. The arrow marks the position of the laser for the pump-probe experiment. (b) Phase shift as a function of applied continuous wave RF. Sharp resonances correspond to the NMR conditions of different isotopes at $B=1$\,T at $T=1.5\, $K. The line is a guide to the eye.}
  \label{fig:1}
\end{figure}

Experiments were performed with a pump-probe technique using a picosecond Ti:Sapphire laser with 1.5\,ps pulse duration operating at a repetition rate of $75.75$\,MHz. The polarization of the pump beam was modulated between $\sigma^+$ and $\sigma^-$ using a photoelastic modulator at a frequency of $50$\,kHz. The probe was linearly polarized and overlapped with the pump at the sample. Both beams were focused to spots with similar diameters of about 40\,$\mu$m and had equal laser powers of 0.6\,mW (power density 48\,W/cm$^2$). The photon energy was fixed at $1.5986\,$eV (wavelength 775.6\,nm), which corresponds to the trion transition, see the black arrow in the inset to the right in Fig.~\ref{fig:1}(a). After passing through the sample, the Faraday ellipticity (FE) of the probe was tested using an optical bridge consisting of a quarter-wave plate, a Wollaston prism, and a balanced photoreceiver. The FE was measured using a lock-in amplifier at the modulator frequency as a function of the path difference between the pump and probe beams. We use the FE instead of the commonly used Faraday rotation, as close to the resonance the spectroscopic response for the ellipticity is stronger~\cite{YugovaFEvsFR09}.

The studied sample (\#031901C) consists of a CdTe/Cd$_{0.78}$Mg$_{0.22}$Te QW heterostructure grown by molecular-beam epitaxy on a (100) oriented GaAs substrate and separated from it by a CdTe/Cd$_{0.78}$Mg$_{0.22}$Te superlattice grown on a thick 2\,$\mu$m Cd$_{0.78}$Mg$_{0.22}$Te buffer layer. The heterostructure has five periods, each of them consists of a 20-nm-thick CdTe QW and a 110-nm-thick Cd$_{0.78}$Mg$_{0.22}$Te barrier. An additional 110-nm-thick barrier was grown on top of this layer sequence to reduce the contribution of surface charges. The barriers were modulation doped with iodine donors. Electrons from the barrier donors, being collected in the QWs, provide a two-dimensional electron gas with a density of about $n_{\text{e}} = 1.1 \times 10^{10}$\,cm$^{-2}$. Weak localization of the resident electrons at the interface fluctuations leads to the appearance of a trion line in the emission spectra, see inset to Fig.~\ref{fig:1}(a). The trion states (T) are formed by a resident electron and an optically created exciton, and are energetically shifted to the lower energy by the binding energy from the free excition transition (X), the higher energy PL line~\cite{ZhukovPRB07}. To facilitate the transmission experiments, the GaAs substrate was chemically removed. During measurements, the sample was mounted inside a bath cryostat at temperature $T = 1.5$\,K. The $g$ factor of the resident electrons $g_e = - 1.64 \pm 0.02$ was determined from the Larmor precession frequency in an external magnetic field of $1\,$T~\cite{Sirenko97,ZhukovPRB07}. Detailed information on the coherent electron spin properties in this sample was published in Refs.~[\onlinecite{Astakhov08,ZhukovPRB07,ZhukovPSSb}]. Extended information on the electron-nuclear spin interaction and the polarization of nuclear spins by optical excitation can be found in Ref.~[\onlinecite{Zhu2014}]. This reference compares several systems with different localization volumes of the resident electrons and considers a case of ODNMR, which is quite different from the presented case, where the nuclear polarization is studied far from the NMR.

To directly manipulate the nuclei, the sample holder was equipped with a copper coil of about $1\,$mm in diameter made out of 10 turns of wire. For the sake of broadband operation, the coil was directly connected to a coaxial cable without impedance matching. It was driven by an RF signal synthesized by a function generator and routed through a $100\,$W pulsed RF amplifier. The coil was placed directly on the sample surface with its opening oriented along the optical axis to apply an RF field with its magnetic component orthogonal to the external static magnetic field. The sample was excited through the opening in the center of the coil.

Figure~\ref{fig:1} shows the typical spin dynamics at $B=1$\,T. The pump pulse at zero delay induces a net electron spin polarization along the optical axis $z$ parallel to the sample growth axis. In a transverse magnetic field $B_x$ the electron spins undergo a Larmor precession with the frequency $\omega_e = g_e \mu_B B_x/\hbar \approx 114.5$\,rad$/$ns at $B=1 \,$T, where $\mu_B$ is the Bohr magneton and $\hbar$ the reduced Planck constant. The precession is observed as oscillatory changes of the ellipticity signal. The amplitude of these oscillations decays due to inhomogeneous dephasing of the electron spin ensemble caused by the electron $g$-factor dispersion~\cite{ZhukovPRB07}.

Generally, the observed Larmor frequency is defined by the common action of the external magnetic field and the Overhauser field of nuclear spins. Their contribution is evidenced by measuring the pump-probe spectra in the presence of an RF field. Once the RF frequency $f_{\text{RF}}$ matches the resonance conditions $f_{\text{RF}} = f_{\text{r}}$ for a specific isotope with a resonance frequency $f_{\text{r}}$, the nuclear spins become depolarized and the Overhauser field reduced. The result of such an experiment is plotted in Fig.~\ref{fig:1}(a) where the red curve is taken under RF field radiation and the blue curve without it. The main difference is a sizable shift of the electron precession frequency (left inset in Fig.~\ref{fig:1}(a)), which indicates that without the RF the nuclear spins were partially polarized along the external magnetic field, affecting the electron spin precession. We evaluate the frequency shift by taking a short excerpt of the spectrum roughly $11\,$ns after the pump-probe coincidence and determine the phase shift of the oscillation with RF in comparison to the case without RF, see the left inset in Fig.~\ref{fig:1}(a). We use this phase shift as an indicator to measure the effect of the RF on the nuclear spin system.

Monitoring the frequency (or equivalently the phase of the oscillation for a given part of the pump-probe delay) of the electron Larmor precession allows us to assess the contributions of individual isotopes separately. The phase of the oscillation was determined by performing a short pump-probe scan around -1.7\,ns covering the range of a single electron-spin precession period at a fixed magnetic field. Figure~\ref{fig:1}(b) shows the resonances of several nuclear isotopes measured by applying a continuous sweep of RF at $B=1$\,T. For this measurement, the function generator was connected directly to the copper coil without using the amplifier. The resonance frequency $f_{\text{r}}$ for a NMR can be calculated as $f_{\text{r}} = |\gamma| B/(2\pi)$. The values of the gyromagnetic ratio $\gamma$ for the $^{111}$Cd ($-56.9$\,[T$\mu$s]$^{-1}$), $^{113}$Cd ($-59.6$\,[T$\mu$s]$^{-1}$) and $^{125}$Te isotopes ($-85.1$\,[T$\mu$s]$^{-1}$) lead to: $f_{\text{r}}$\,[MHz] $= 9.06 B$\,[T], $9.48 B$\,[T] and $13.52 B$\,[T], respectively~\cite{PhysHandbook}. Negative values of $\gamma$ are related to the direction of the nuclear field precession with respect to the external field. As seen in Fig.~\ref{fig:1}(a), for the negative sign, the Overhauser field is oriented anti-parallel to the external field, and the Larmor frequency is reduced for polarized nuclei (i.e., the signal without RF).

The asymmetry of the peaks in Fig.~\ref{fig:1}(b) is a result of a too short time interval between subsequent RF steps, which did not allow the nuclei to reach a steady-state polarization, see below for the measurement of the longitudinal spin relaxation time $T_1$. The direction of the scan was from low to high frequencies. It is important to mention, that all isotopes of CdTe have nuclear spin $I=1/2$, so no quadrupole effects have to be considered. This simplifies the interpretation of the results and leads to a strong spectroscopic response, as all nuclei of the same isotope have the same Zeeman splitting.

In the experiments presented below, we fix the magnetic field at $1\,$T, the RF at resonant frequencies, and apply RF pulses. The following measurement cycle is used: The laser pulses polarize the electrons for five minutes for the $T_1$ measurements or forty seconds for all other measurements. Then we choose one of two scenarios: (i) apply RF and continue to apply the laser radiation, or (ii) apply RF, but block the laser radiation during RF application. Afterward, the laser radiation is unblocked, and the phase shift of the electron precession is read out with the RF switched off. Then the cycle is repeated for the next data point.

\begin{figure}[th]
  \includegraphics[scale=2.7]{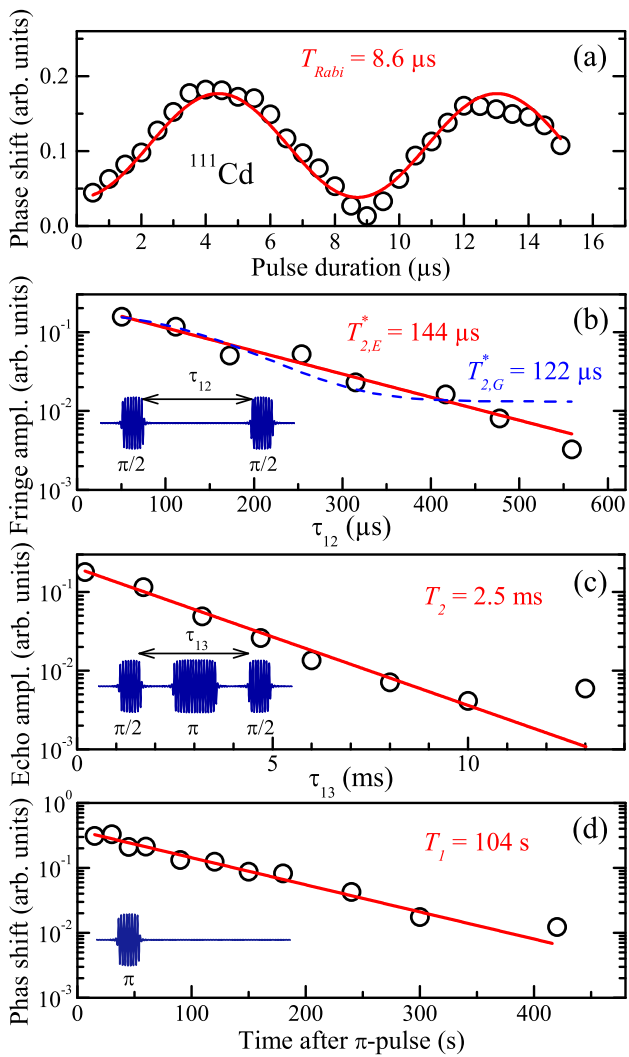}
  \caption{Coherent control of the $^{111}$Cd isotope at $B=1$\,T and $f_{\text{r}}=9.06$\,MHz. (a) Phase shift of the electron precession induced by an RF pulse. The open circles correspond to a 650\,mV pulse with varied duration and the red curve is a sine fit with a Rabi period of $8.6\,\mu$s. (b) Decay of the Ramsey fringes amplitude. The red line corresponds to an exponential decay with a time constant of 144\,$\mu$s, whereas the blue dashed curve corresponds to Gaussian fit with a decay constant of 122\,$\mu$s. (c) Decay of the spin-echo amplitude. The line is an exponential decay fit featuring a time constant of 2.5\,ms. (d) Decay of the phase shift with recurring nuclear polarization. The line is an exponential decay fit with a time constant of 104\,s. The measurements in panels (b)-(d) are given for RF application while the laser radiation was blocked. The sketches show the corresponding pulse sequences used in the experiments.}
  \label{fig:2}
\end{figure}

Resonant RF pulses allow us to provide a coherent nuclear spin control of specific isotopes, where the action of the RF can be considered as an action of an effective magnetic field along the $z$-axis, which produces a coherent rotation of the nuclear spin in the $xy$-plane. The dependence of the electron precession phase shift on the RF pulse parameters is presented in Fig.~\ref{fig:2}(a) for the $^{111}$Cd isotope at $f_{\text{r}}=9.06$\,MHz. Clear oscillatory behavior is observed as a function of the pulse duration, which is an unequivocal signature of Rabi oscillations of the addressed nuclear spins. The period of oscillations allows us to determine the pulse duration corresponding to $\pi/2$ and $\pi$ rotations and to characterize the value of the effective magnetic field produced by the RF-coil. So, for a full $2\pi$ Rabi period with an applied voltage of $650$\,mV, it requires a 8.6\,$\mu$s pulse, which corresponds to $B_{\textrm{eff}}=2\pi/$(8.6\,[$\mu$s] 0.0569\,[mT$\mu$s]$^{-1}$)$=12.8$\,mT.

\section{Results}
Figure~\ref{fig:2}(b) demonstrates the Ramsey fringe experiment~\cite{Ramsey1950} and the corresponding inhomogeneous spin dephasing time $T_2^*$ of the $^{111}$Cd isotope. The sequence for this experiment consists of two $\pi/2$ pulses separated by a delay $\tau_{12}$. The first pulse rotates the nuclear spins out of the $x$-axis to the $y$-axis, where they start to precess around the $x$-axis in $yz$-plane. The second pulse brings the nuclear spins back to the $x$-axis. Depending on the phase accumulated over the time $\tau_{12}$, the maximally achieved spin polarization of the ensemble decreases. Assuming an exponential decay of the oscillation amplitude, we fit the value of $T_{2,\text{E}}^* = 144\,\mu$s for the studied $^{111}$Cd isotope (red curve in Fig.~\ref{fig:1}(b)). A Gaussian decay delivers a dephasing time of $T_{2,\text{G}}^* = 122\,\mu$s (blue dashed curve in Fig.~\ref{fig:2}(b)). Based on the results, the exponential fit gives a better description.

Figure~\ref{fig:2}(c) shows the next sequence, the spin Hahn-echo~\cite{Hahn1950}. The first and the third pulse have an area of $\pi/2$ and serves for the same purpose as in the previous case. The first pulse rotates the nuclear polarization by $\pi/2$ to the $yz$-plane, and the third pulse rotates it back to the original orientation along the $x$-axis. The second pulse of that sequence has an area of $\pi$ and is delayed by approximately $\tau_{13}/2$ after the first pulse. The role of the second pulse is to invert the nuclear polarization by~$180^{\circ}$, which reverses the dephasing of the ensemble and allows to cancel the effect of the Larmor frequency spread. By comparing the effect of the first (Fig.~\ref{fig:2}(b)) and the second sequence (Fig.~\ref{fig:2}(c)) we see that the introduction of the $\pi$ pulse considerably extends the timescale over which a coherent nuclear precession is observed. In that case, we extract a coherence time of $T_2 = 2.5$\,ms using an exponential fit.

To complete the characterization of the nuclear spin dynamics, we provide the measurement of the longitudinal nuclear spin relaxation time $T_1$. Here, a single $\pi$-pulse is used to invert the nuclear spin polarization along the $x$-axis after initial optical pumping of the nuclear system. It causes a sudden phase shift of the electron precession followed by a slow recovery on the timescale of $T_1$. Figure~\ref{fig:2}(d) shows the corresponding measurement which exhibits a longitudinal spin relaxation time of $T_1 = 104$\,s. It demonstrates that the condition $T_1 \gg T_2$ is valid in this sample, and the spin temperature of the nuclei reaches its equilibrium much faster than the energy transfers to the lattice. Furthermore, this experiment allows us to extract the degree of nuclear polarization from the phase shift of the electron precession right after the application of the RF pulse. We determine the change of the nuclear field $B_{\text{N}}$ acting on the electron spin to be $1.5\,$mT when manipulating $^{111}$Cd isotopes by a single $\pi$-pulse. This value corresponds to $6\,\%$ of the maximal nuclear spin polarization of $^{111}$Cd, which can be estimated as $B_{\text{N}}/B_{N,max}= 1.5\,\text{[mT]}/25.25\,\text{[mT]}=0.06$, see Ref.~[\onlinecite{Zhu2014}]. The field $B_{\text{N,max}}$ is the maximal field induced by a completely polarized nuclear spin of $^{111}$Cd~[\onlinecite{Zhu2014}]. Other isotopes give similar polarization values: 5.5\% for $^{113}$Cd and 9.7\% for $^{125}$Te. All measurements presented in Figs.~\ref{fig:2}(b)-(d) were realized with the second scenario, where the RF pulses were applied, while no optical excitation of the sample was present. These measurements were performed for $^{111}$Cd, $^{113}$Cd, and $^{125}$Te and the times are summarized in Tab.~\ref{tab:times}.

\begin{table}
  \centering
  \caption{Relaxation times for all nonzero spin isotope species in the CdTe QW. Values are given without and with illumination, corresponding to the nuclear spin dynamics in scenario (ii)/scenario (i).}
  \begin{tabular}{crrr}
    Isotope & $T_2^*$\,[$\mu$s] & $T_2$\,[ms] & $T_1$\,[s] \\
    \hline
    $^{111}$Cd & 144/366 & 2.5/2.9 & 104/173 \\
    $^{113}$Cd & 191/446 & 2.6/4.6 & 116/169 \\
    $^{125}$Te & 200/200 & 2.6/3.8 & 130/138 \\
  \end{tabular}
  \label{tab:times}
\end{table}

One noticeable feature of the times of the different isotopes measured in the dark condition (without laser radiation) is their similarity to each other. It cannot be related to a nuclear spin diffusion, which happens due to a dipole-dipole interaction of the nuclei of a particular isotope. In the case of CdTe QWs, the electrons are weakly localized and, therefore, do not produce a strong inhomogeneity of the Knight field. Correspondingly, the nuclear field distribution only shows a small inhomogeneity which weakly affects the nuclear spin diffusion~\cite{Heist15}. Furthermore, all nonzero nuclear spin isotopes have spin 1/2, which simplifies the spin diffusion due to the absence of quadrupole splittings. However, at a magnetic field of 1\,T, the spin diffusion can happen purely within one type of isotopes, as the difference in the Zeeman splitting of different isotopes is much bigger than the local field of the nuclei. Another option to couple different types of isotopes and nuclei of one isotope with each other is the possibility to use the electron spin as a mediator in a flip-flop process of two nuclear spins. This process is shown to be efficient in quantum dots~\cite{Malet07,Warb16} and can be applied to the present case as well.

The previously described measurements were done without sample illumination during the application of RF fields. We repeated them with ongoing optical electron-spin orientation during the RF sequences. These experiments were performed for all involved times. The times are also presented in Tab.~\ref{tab:times} after each slash. In all experiments, the presence of a pump-probe excitation during RF action does not lead to a shortening of the nuclear spin relaxation times. On the contrary, some times are prolonged under illumination. This is a striking result. Using the results discussed in Ref.~[\onlinecite{Warb16}], we can assume that the $T_{1,e}$ of electrons is anticorrelated with the $T_2$ of nuclei. For higher excitation power, the spin lifetime ($T_{1,e}$) of electrons becomes reduced (see e.g. Ref.~[\onlinecite{Heist15}] for ZnSe:F or Ref.~[\onlinecite{Zhukov18}] for (In,Ga)As/GaAs quantum dots), and therefore one could expect an increase of the nuclear $T_2$ time, which is consistent with our results. The changes of $T_2$ of nuclei in our case are not as dramatic as in quantum dots, due to a much weaker localization of electrons and therefore reduced electron-nuclear interaction, see Ref.~[\onlinecite{Warb16}]. Furthermore, Refs.~[\onlinecite{Greilich2007,Warb16}] report on a direct correlation between $T_{1,e}$ of electron spins and $T_1$ of nuclei. Similar dependence was observed in Ref.~[\onlinecite{Heist16}], where the authors compared the longitudinal polarization dynamics in the dark scenario to the dynamics in the bright scenario for a system with a strongly localized donor electrons and reported a reduction of the nuclear $T_1$ time by three orders of magnitude under illumination. The result, that in our case, the nuclei $T_1$ time slightly increases under the illumination is still puzzling to us and requires further investigations.

\section{Conclusions}
To conclude, we present a comprehensive characterization of the nuclear spin dynamics in CdTe/(Cd,Mg)Te QWs. We use ODNMR based technique on a pump-probe setup and manipulate the nuclear polarization directly by using RF pulses. The characteristic time scales are similar between all isotopes and demonstrate a weak stabilizing effect of the laser illumination.

\section{Acknowledgements}
We acknowledge the financial support by Deutsche Forschungsgemeinschaft in the frame of the ICRC TRR 160 (Projects A1 and A6) and Russian Science Foundation (Grant No. 14-42-00015).

  \bibliographystyle{apsrev}

\end{document}